\documentclass[twocolumn]{aastex701}
\usepackage{xcolor}
\usepackage[T1]{fontenc}
\usepackage{subfigure}
\usepackage{hyperref}
\usepackage{longtable}
\usepackage[utf8]{inputenc}
\usepackage{pifont}
\usepackage{amsmath}
\usepackage{stmaryrd}
\usepackage{amssymb}
\usepackage{wasysym}
\usepackage{amsbsy}
\usepackage{soul}
\usepackage{xcolor}

\DeclareUnicodeCharacter{2212}{-}

\begin{document}
\title{Revisiting neutrino event epochs for the blazar PKS 0735+178 with TESS}
\author[0000-0001-8716-9412]{Shubham Kishore}
\affiliation{Indian Institute of Astrophysics, Bangalore (IIA), 560034, India}
\email[show]{amp700151@gmail.com}  
\author[0000-0002-9331-4388]{Alok C.\ Gupta}
\affiliation{Aryabhatta Research Institute of Observational Sciences (ARIES), Manora Peak, Nainital 263001, India}
\email{acgupta30@google.com}
\author[0000-0003-1071-5854]{Debanjan Bose}
\affiliation{Central University of Kashmir, Ganderbal, 191131, India}
\email{debaice@gmail.com}
\begin{abstract}
\noindent
We present here the results of the optical light curve variability analysis of the blazars PKS~0735+178, in weeks-scale flare state, observed in three sectors with the Transiting Exoplanet Survey Satellite (TESS). The TESS observations in this study coincide with a well-known neutrino emission phase detected with four different neutrino observatories at multiple epochs in a narrow time window. We segmented the rising and decaying parts of the flare and individually analyzed their flux distribution, excess variance, variability timescale, and the power spectral density (PSD). The source displayed an elevated excess variance of $\sim25\%$, with a multi-modal flux distribution (coherent in the rising and distorted in the decaying phase). The variability timescale analysis highlights a much faster decay than the rising scale, and the PSDs depict a nominal change in the power spectral slope. We discuss a likely connection in the optical variations and the neutrino events, and briefly explain a possible physical scenario for the observed optical flux behavior in view of previously discovered radio-band results.
\end{abstract} 
\keywords{\uat{Galaxies}{573} --- \uat{Active Galactic Nuclei}{16} --- \uat{Blazars}{164} --- \uat{Optical Astronomy}{1776} --- \uat{High Energy Astrophysics}{739}}
\section{Introduction}\label{sec:intro}
\noindent
Flux variability is one of the defining characteristics of active galactic nuclei (AGNs), which are understood to possess accreting supermassive black holes (SMBHs) with masses in the range of 10$^{6} - \rm{10}^{10} \rm{M}_{\odot}$.  Blazars are a subclass of radio loud AGNs that launch powerful, large-scale relativistic jets of plasma pointed towards the line of sight (LOS) ($\leq \rm{10}^{\circ}$) of the observer \citep[e.g.][]{1995PASP..107..803U}.  BL Lacertae objects (BL Lacs) and flat-spectrum radio quasars (FSRQs) are jointly called blazars. In the composite ultraviolet (UV) to optical spectra, BL Lacs possess either an absence or very weak narrow emission lines (equivalent width, EW $\leq$~5\AA), whereas FSRQs show broad and strong emission lines. \\
\\
Blazars display large flux and spectral variability on diverse timescales throughout the observable electromagnetic (EM) bands and show substantial polarization that is also often variable in the radio, optical, and X-ray bands \citep[e.g.,][and references therein]{2018ApJS..234...12L, 2019AJ....157...95G, 2024ApJ...972...74M}. Blazar variability having timescales of a few minutes to less than a day is often called microvariability \citep{1989Natur.337..627M} or intraday variability (IDV) \citep{1995ARA&A..33..163W} or intranight variability \citep{1995MNRAS.274..701G}. Variability occurring over a few days to weeks is known as short-term variability (STV), while that occurring over periods longer than months is called long-term variability (LTV) \citep{2004A&A...422..505G}. The nature of blazar light curves (LCs) in the whole EM spectrum is mostly non-linear, stochastic, and aperiodic \citep[e.g.][]{2017ApJ...849..138K}.  \\
\\
The emission from blazars in the whole EM spectrum is predominantly nonthermal. The spectral energy distributions (SEDs) of blazars display a characteristic double-humped structure in the $log~\nu~\text{versus}~log~\nu F_\nu$ diagram \citep{1998MNRAS.299..433F}. The lower-energy hump lies in the radio to UV or, sometimes, soft X-rays, most likely due to polarized Synchrotron emission from relativistic electrons in the hot plasma in a magnetic environment. Depending on the location of the peak of this lower energy hump in the mentioned EM regime, the blazars are often termed as low-, intermediate-, and high- Synchrotron peaked (LSP: $\nu_{peak}<10^{14}Hz$, ISP: $10^{14}Hz<\nu_{peak}<10^{15}Hz$, and HSP: $\nu_{peak}>10^{15}Hz$) blazars \citep{2010ApJ...716...30A}. The higher-energy hump covers X-rays to $\gamma-$rays, which is most likely due to inverse Compton emission from the interaction of relativistic electrons (under the so-called `leptonic model') and the low-energy seed photons. When the seed photons originate from the Synchrotron emission itself, the mechanism is termed Synchrotron Self Compton (SSC); otherwise, it is called the External Compton (EC) mechanism.\\
\\
% \textcolor{red}{Apart from the leptonic origin, the higher-energy humps are sometimes also believed to be caused by the Synchrotron emission from relativistic protons in magnetized media}.
Beyond leptonic processes, the high-energy component is often attributed to hadronic interactions. In this scenario, gamma-rays can be produced from the decay of neutral pions ($\pi^0$) in proton-photon or proton-proton collisions \citep{1993A&A...269...67M}. Alternatively, these high-energy $\gamma$-rays may be generated through synchrotron emission from relativistic protons in magnetized media \citep{2000NewA....5..377A}. 
A new perspective on the non-thermal processes in our Universe was made possible by the IceCube Neutrino Observatory's 2013 discovery of high-energy neutrinos of cosmic origin \citep{2018AdSpR..62.2902A}. The origin of ultra-high-energy cosmic rays (UHECRs), the Universe's most energetic particles with energies more than 10$^{18}$~eV, has long been a mystery \citep{2017PTEP.2017lA101D}. A largely extragalactic origin is supported by the observed distribution of their arrival directions in the sky \citep{2017Sci...357.1266P}. It was predicted more than one and a half decades ago that neutrino emission can be detected from some blazars \citep{2002PhRvD..66l3003N}. On 22 September 2017, the IceCube neutrino observatory detected the first likely neutrino event from the blazar TXS 0506+056 at energy $\sim$290~TeV, which was identified as IceCube-170922A \citep{2018Sci...361..147I,2018ApJ...863L..10A}. The prerequisites for neutrino emission from blazars are through proton-photon interactions, i.e., a situation where the shock may accelerate protons to relativistic energies, with the target photons  \citep{2025ApJ...989..208P}. Parsec-scale relativistic jets play an important role in neutrino production in blazars. The relativistic protons present in the jet collide with the low-energy photons (optical/UV/x-ray) present in the surrounding ambient medium and produce pions. The decay of charged pions ultimately produces high-energy neutrinos. A detailed model and explanation is given in \citet{2021ApJ...908..157P}.\\
\\
PKS~0735+178 at z = 0.45$\pm$0.06 \citep{2012A&A...547A...1N}, was first optically identified by \citet{1970ApL.....6..201B} and classified as a BL Lac \citep{1974ApJ...190L.101C}. The source is optically bright, its host galaxy is unresolved in optical imaging, and the optical spectrum is a featureless continuum \citep{1974ApJ...190L.101C}.  The source has been extensively studied in the optical bands in the search for flux variability on the diverse timescales which were often detected \citep[e.g.,][and references therein]{2007A&A...467..465C, 2008AJ....135.1384G, 2009MNRAS.399.1622G, 2017ApJ...837..127G, 2022ApJ...933..224F}. The source has displayed a high degree of optical and near-infrared polarization upto 35\% \citep[e.g.,][and references therein]{1990A&AS...83..183M, 1991A&AS...90..161T, 1992A&AS...94...37T, 1991AJ....101...78V, 1993A&A...278..371V, 2001A&A...376...51T, 2011ApJS..194...19W}. There are detections of periodic and quasiperiodic oscillations on diverse time scales in optical LCs of the source \citep{1997A&AS..125..525F, 2004PASP..116..161Q, 2007A&A...467..465C, 2026MNRAS.tmp.1208K}. Multi-wavelength variability, cross-correlation in different EM bands, and SEDs are also studied for the source \citep{2017ApJ...837..127G, 2022ApJ...933..224F, 2023MNRAS.519.1396S}.\\
\\
During early December of 2021, the source PKS~0735+178 was found to be in spatial vicinity  \citep[$\sim2.05^{\text{o}}$,][]{2021GCN.31194....1G} with a neutrino event (\href{https://gcn.nasa.gov/circulars/events/icecube-211208a}{IceCube-211208A}) of \(\sim171~\)TeV. The event triggered inspection of the related sky region with other facilities such as with Baikal \citep[43~TeV,][]{2021ATel15112....1D}, Baksan \citep[\(>\)1~GeV,][]{2021ATel15143....1P} and KM3NeT \citep[18TeV,][]{2022ATel15290....1F} while it was undergoing its largest $\gamma-$ray, optical, and soft X-ray flare observed since the launch of the Fermi satellite in 2008 \citep{2023MNRAS.519.1396S}. Preliminary analysis indicates that on 2021-12-08, the source was observed with a daily averaged flux (E\(>\)100 MeV) being about 5 times greater than the average value indicated in the fourth Fermi LAT source catalog (4FGL-DR2), and 
its peak daily flux value was already observed to be about 10 times greater than the average 4FGL-DR2 flux on 2021-12-04. The neutrino event was reported by four observatories during the December 2021 flare exhibited in the source in $\gamma-$ray, X-ray, ultraviolet, infrared, and optical bands \citep{2023ApJ...954...70A}. \citet{2024MNRAS.527.8746P} analyzed the 100 days $\gamma-$ray and X-ray data, which were 10 days around the neutrino event, and generated broad-band SEDs and modeled them by fitting synchrotron and SSC and invoking photohadronic ($p\gamma$) interactions inside the jet. The SED of the source can be well modeled under the pure leptonic emission scenario, according to another MW SED analysis conducted during the neutrino event. This is likely because the hadronic emission contribution is subdominant to the leptonic process \citep{2024MNRAS.529.3503B}. By using very-long-baseline interferometry polarized light images of PKS 0735+178, no new jet ejection component was observed during that narrow time frame of neutrino emission event \citep{2025ApJ...989..208P}. \citet{2026MNRAS.549ag658B} examined long-term VLBA observations of the source, having 40 epochs of 15 GHz data (1995.27–2025.02) and 116 epochs of 43 GHz data (2008.62–2025.22), which also covered the time of the neutrino detections. They proposed that jet precession aligns the more nuclear parts of the jet into loops and rings due to gravitational lensing by a known intervening absorption system (z = 0.424). The MW variability of the source for which light curves span the radio (1–230 GHz), optical, and FERMI-LAT $\gamma-$ray bands about three-decade duration was recently reported. Delays from 0 to 1200 days increase towards lower frequencies in the associated light curves, which is compatible with emission from an opacity-stratified jet \citep{2026MNRAS.550g1179M}.\\
\\
In this paper, we present results of the time-domain analysis of a high-cadence densely sampled LC of the blazar PKS~0735+178 that was fortunately within the field-of-view of the Transiting Exoplanet Survey Satellite (TESS) during the well-recognized neutrino event in early December~2021. No known previous study of this source utilized such a high-cadence sampled dataset over an appreciable duration covering all three cases of pre-, mid-, and post-event, which is helpful in getting the temporal aspects of the source during the neutrino emission epochs.  \\
\\
The paper is structured as follows. In Section 2, we discuss the TESS data acquisition and reduction that we carried out. Section 3, 4, 5, and 6 provide descriptions of the data analysis employed and the results in the present work.  Section 7 includes the discussions and conclusions. 
\begin{figure*}
    % \centering
    \includegraphics[width=1.0\linewidth]{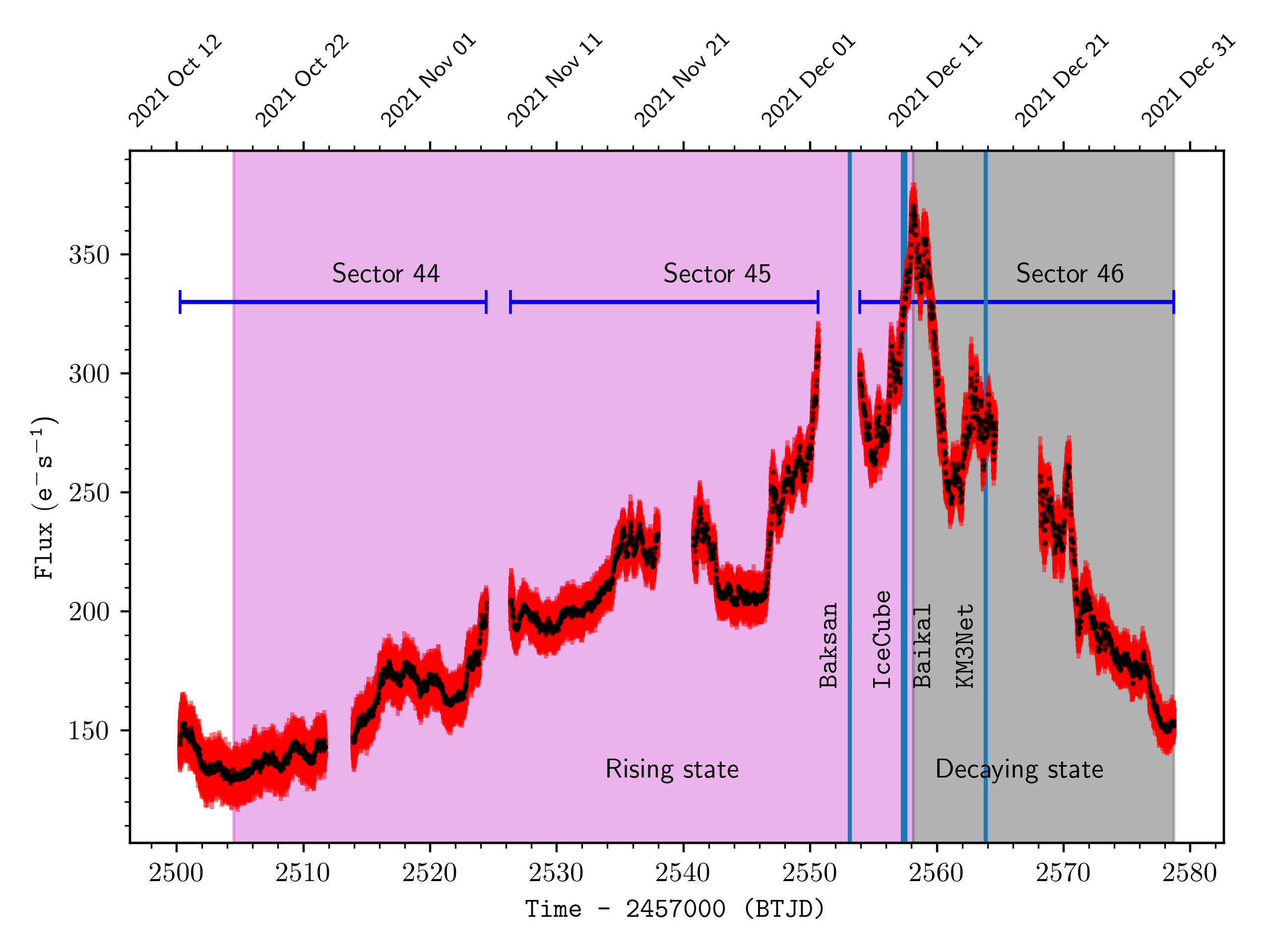}
    \caption{LC of the blazar PKS 0735+178, observed in the three continuous indicated sectors (44, 45, and 46); the red LC is of 2-min cadence, and the black LC is 30-min binned, which have been used in the analyses.   }
    \label{fig:complete_LC}
\end{figure*}
\section{Data Acquisition and Light Curves}
\begin{figure*}
\centering
\includegraphics[width=.475\linewidth]{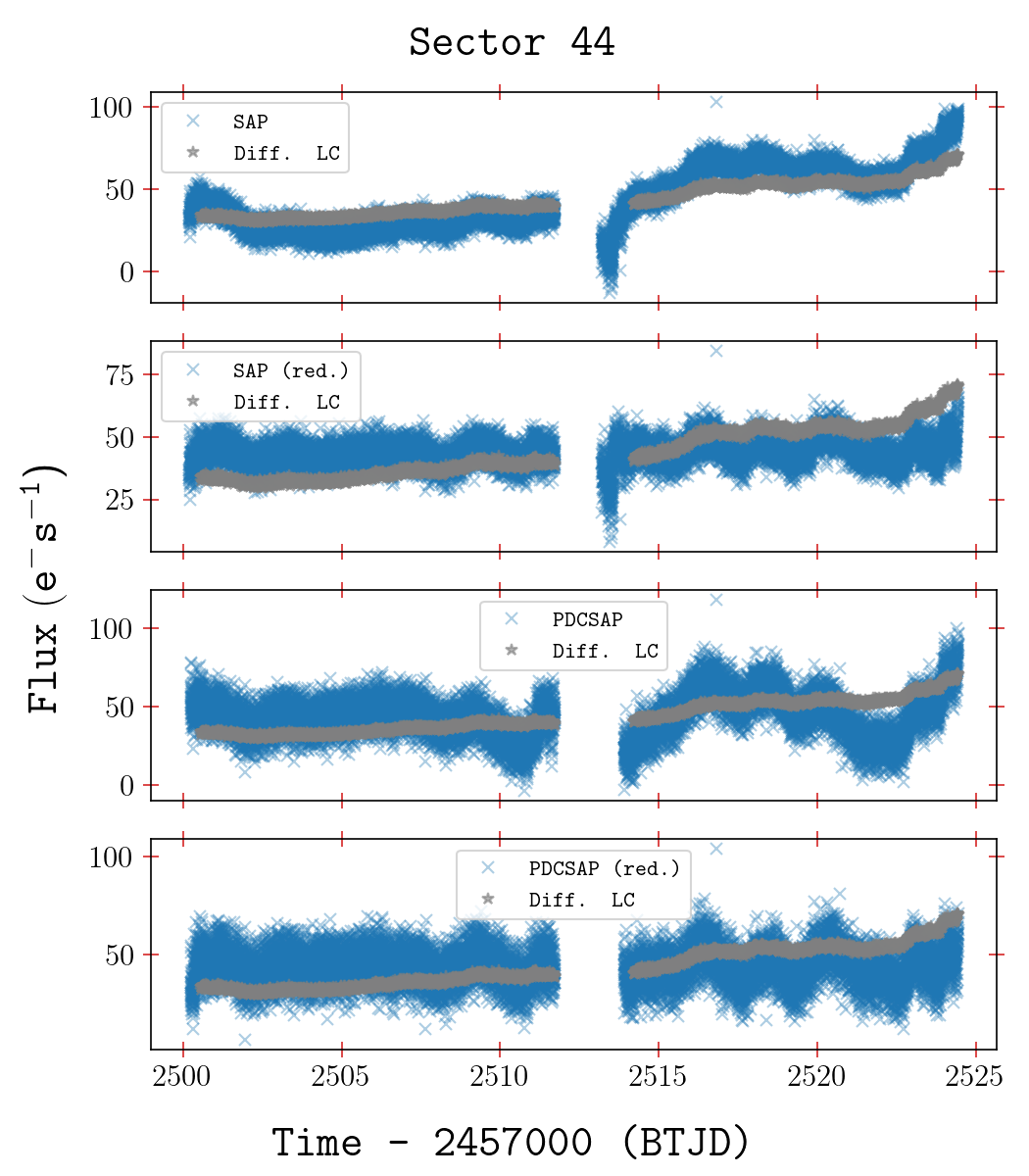}
\includegraphics[width=.475\linewidth]{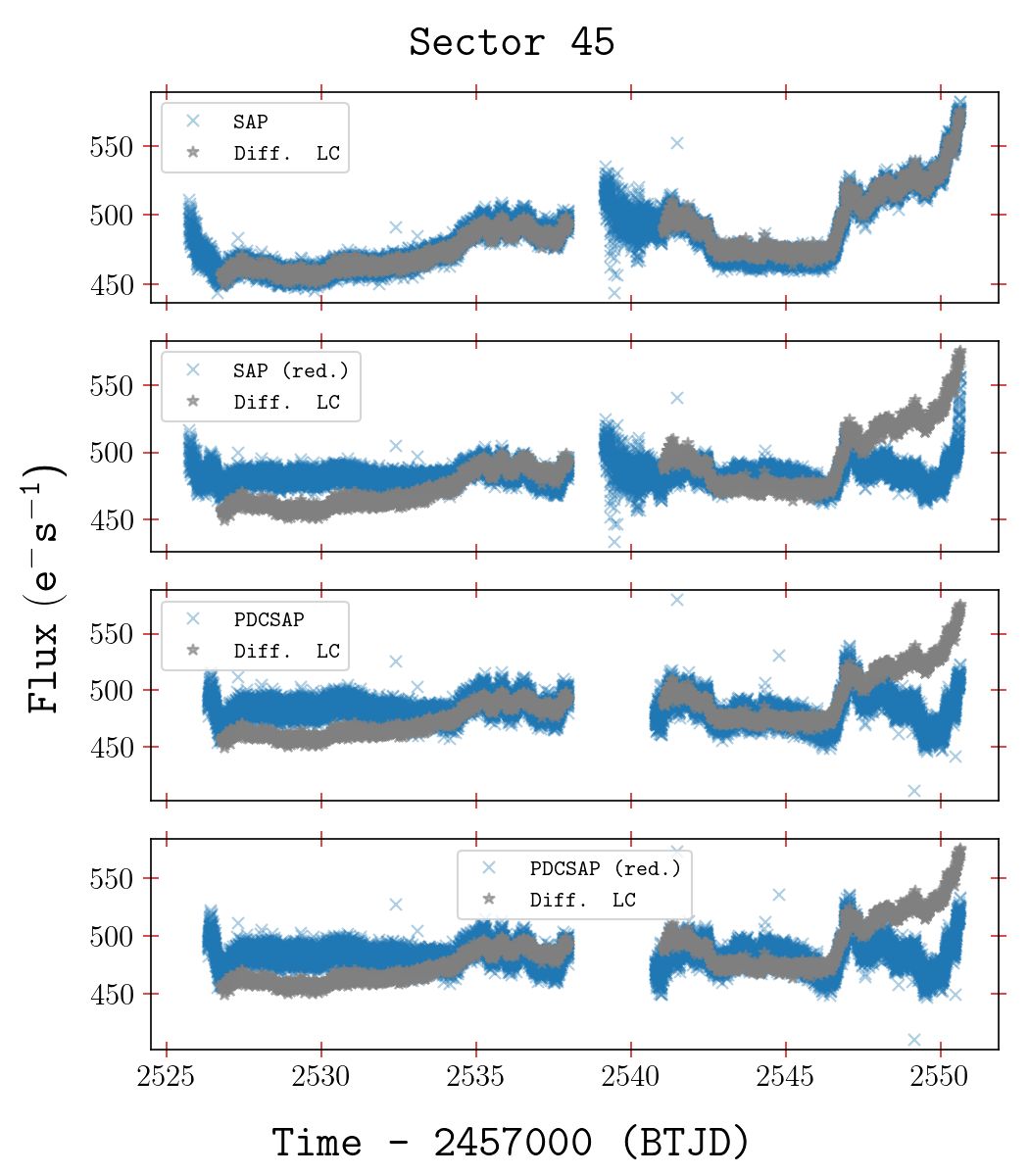}
\includegraphics[width=.48\linewidth]{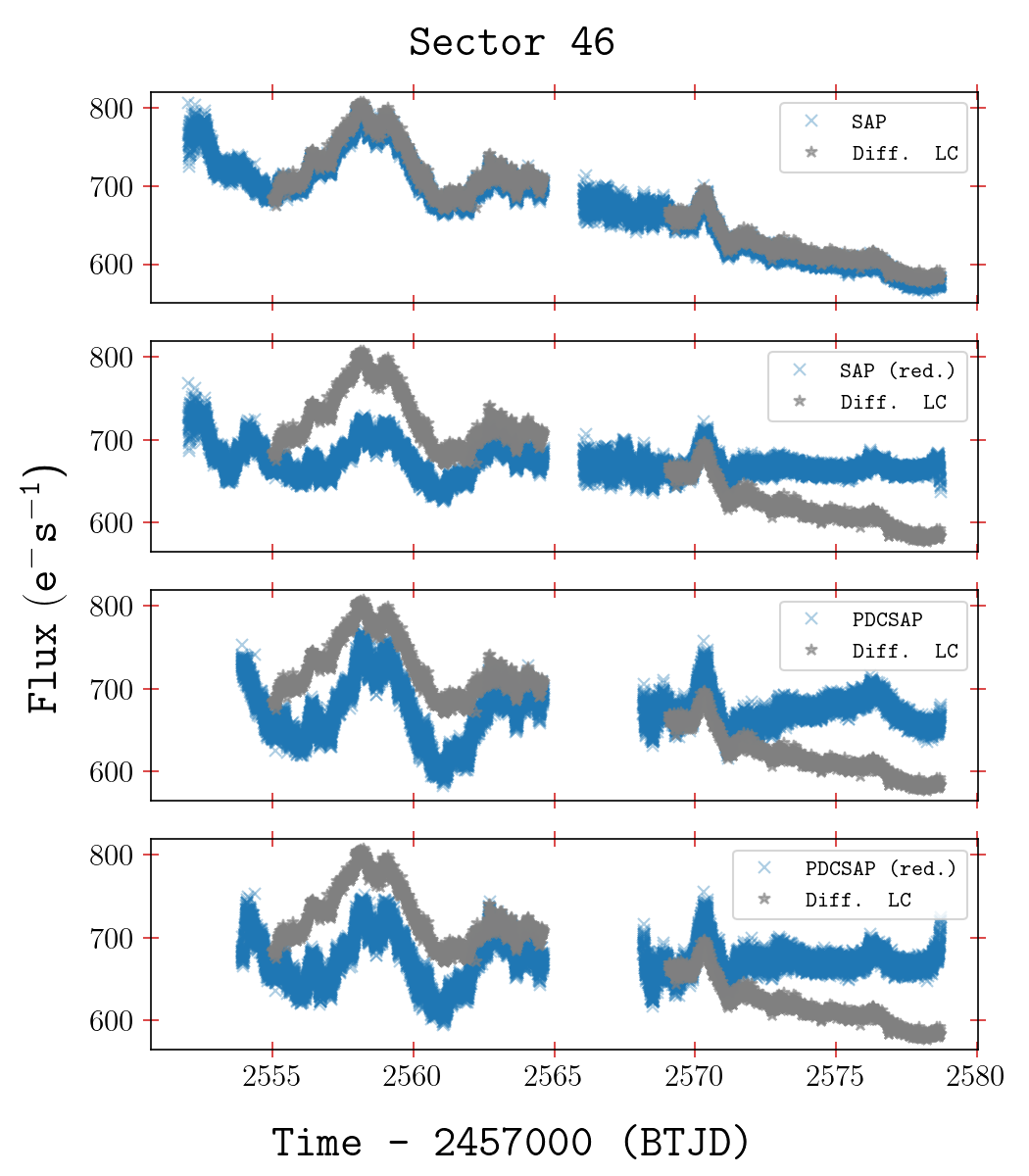}
\caption{Comparison of the differential LC's behavior with those obtained with other methods. The suffix `red.' denotes the corresponding flux from which long-term trending has been reduced. To visualize and compare the LC variations for each of the sectors efficiently, SAP, SAP (red.), PDCSAP and PDCSAP (red.) LCs are shifted along the Y-axis for their mean fluxes to coincide with the mean of the diff. LCs.}
\label{fig:complc}
\end{figure*}

\noindent 
Readers are requested to hover over key papers such as \citet{2014SPIE.9143E..20R, 2016SPIE.9913E..3EJ, twicken2020tess} for the details of instruments used and the data achieved. The details of the TESS instruments are available online\footnote{\href{ https://heasarc.gsfc.nasa.gov/docs/tess/}{https://heasarc.gsfc.nasa.gov/docs/tess/}} and have been briefly discussed in \cite{Kishore_2023} along with the description of two types of fluxes that TESS offers. Also discussed in our previous paper are the parameters that we used in the data reduction: an overfitting metric, an underfitting metric, and a regularization parameter ($\alpha$), along with their optimum values for good data reduction.
Briefly, TESS offers two types of flux values: the SAP\_FLUX (summing all pixel values in a pre-defined aperture corresponding to a source) and its detrended form, PDCSAP\_FLUX. Both of these flux values can be further processed to remove any persisting instrumental artifacts. However, the detrending can majorly affect any genuine flux variation of the source objects, particularly when the sources display stochastic LCs, variable on all timescales, i.e., blazars (or AGNs). Evidence of this effect can be seen as jumps in the mean sectorial flux levels of the source when observed in multiple sectors. In such cases, the LC obtained with the differential photometry can very efficiently guide the appropriateness of the SAP\_FLUX and PDCSAP\_FLUX and their processed versions for a particular sector.
\begin{figure}[h]
    \hspace{-.5cm}
    \includegraphics[width=1.05\linewidth]{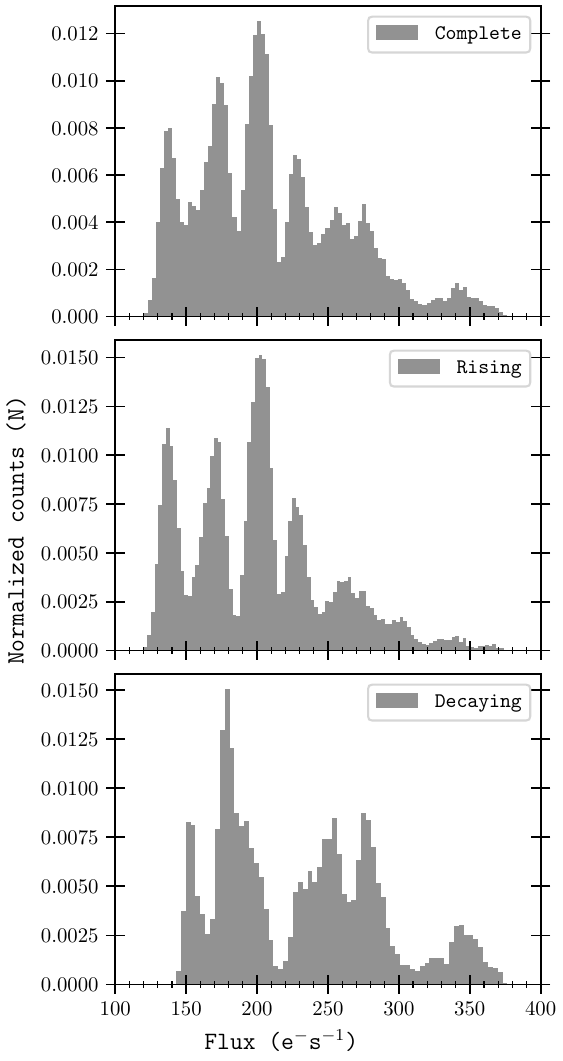}
    \caption{Top, middle, and bottom: Flux distribution of complete flare portion (highlighted in magenta and grey regions in Fig.~\ref{fig:complete_LC}),  rising part (magenta region), and decaying portion (grey region), respectively.}
    \label{fig:flux_dist}
\end{figure}
\begin{figure}[h]
    \centering
    \includegraphics[width=1\linewidth]{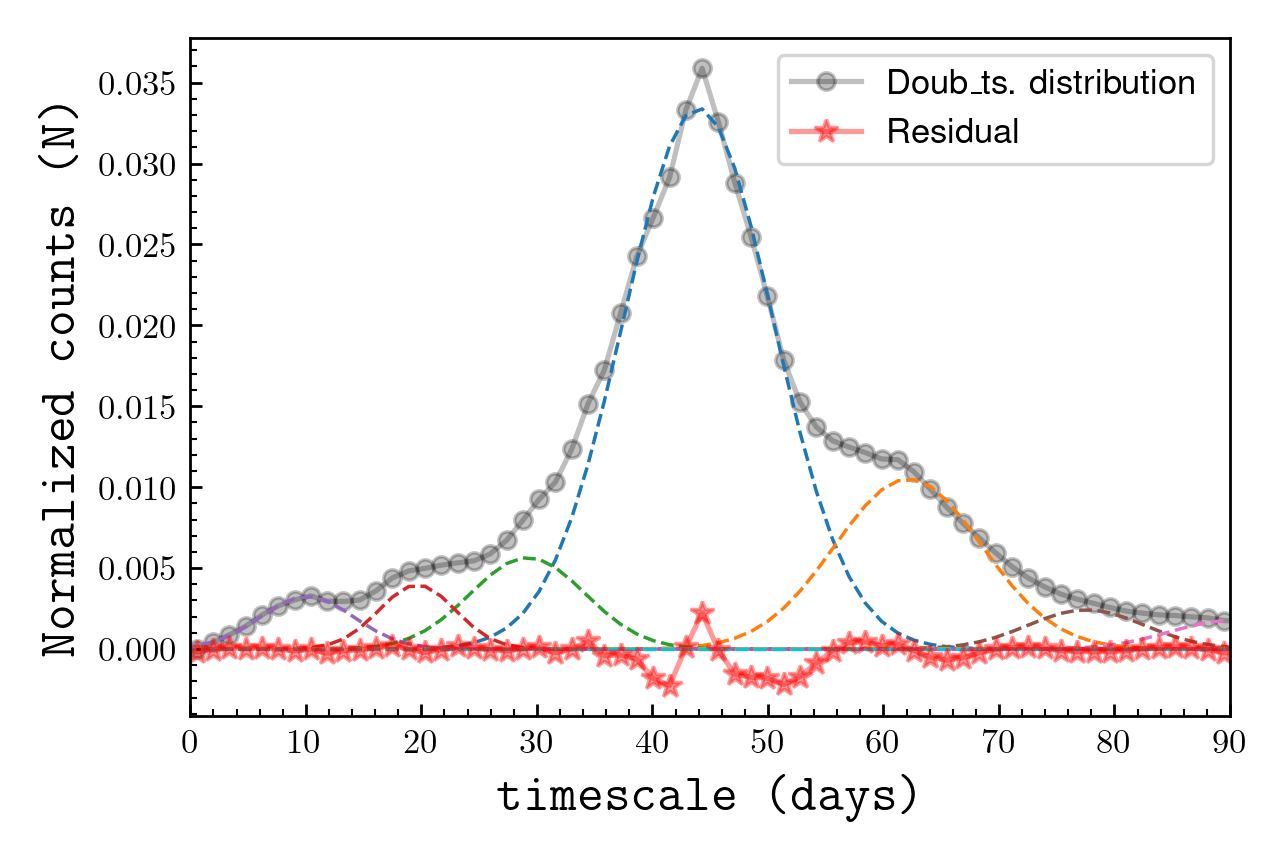}
        \includegraphics[width=1\linewidth]{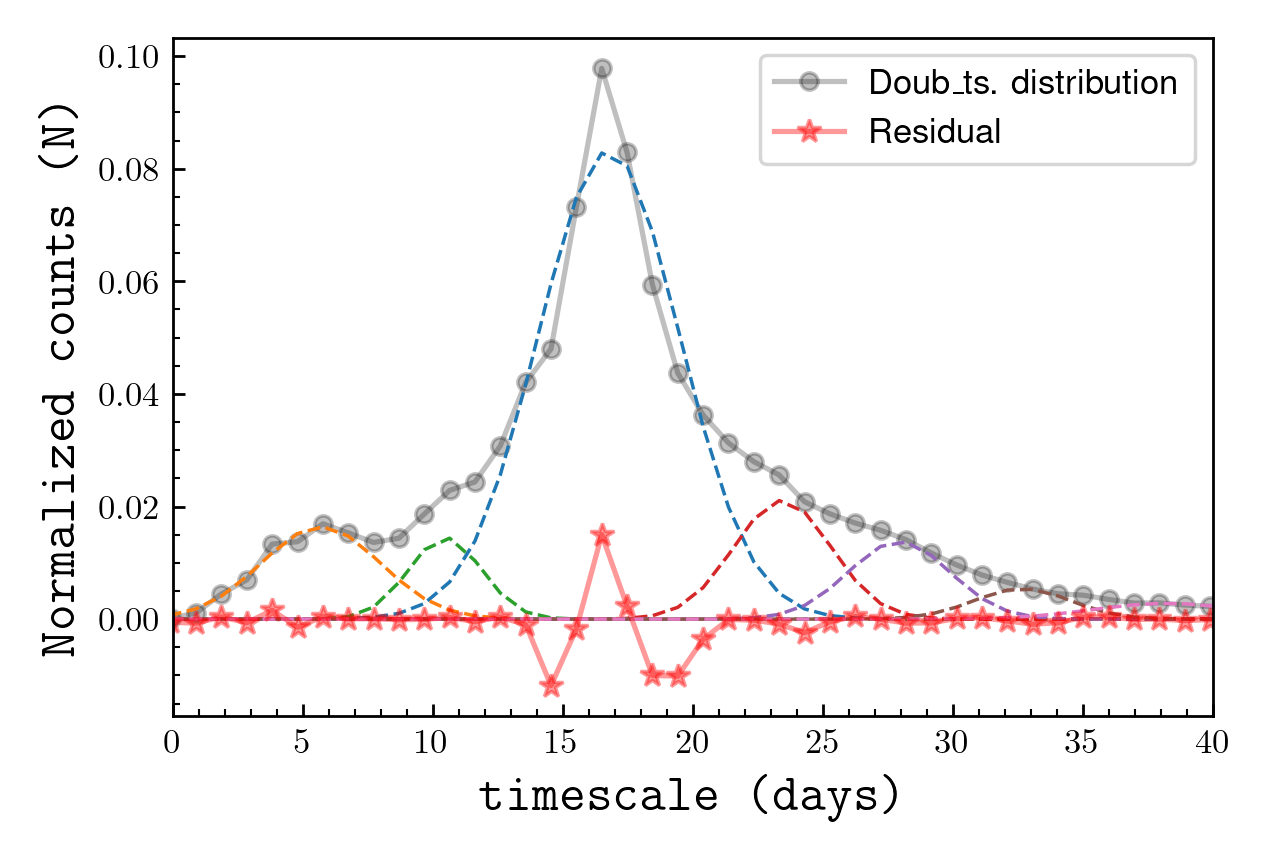}
        \caption{Halving/Doubling timescale distribution for the rising phase (upper panel) and the decaying phase (lower panel); the distributions have been fit using a linear combination of multiple Gaussian profiles to infer the relative contributions of peculiar timescales.}
        \label{fig:doub_ts}
\end{figure}
\begin{figure}[h]
    \centering
    \hspace{-.5cm}
    \includegraphics[width=1.05\linewidth]{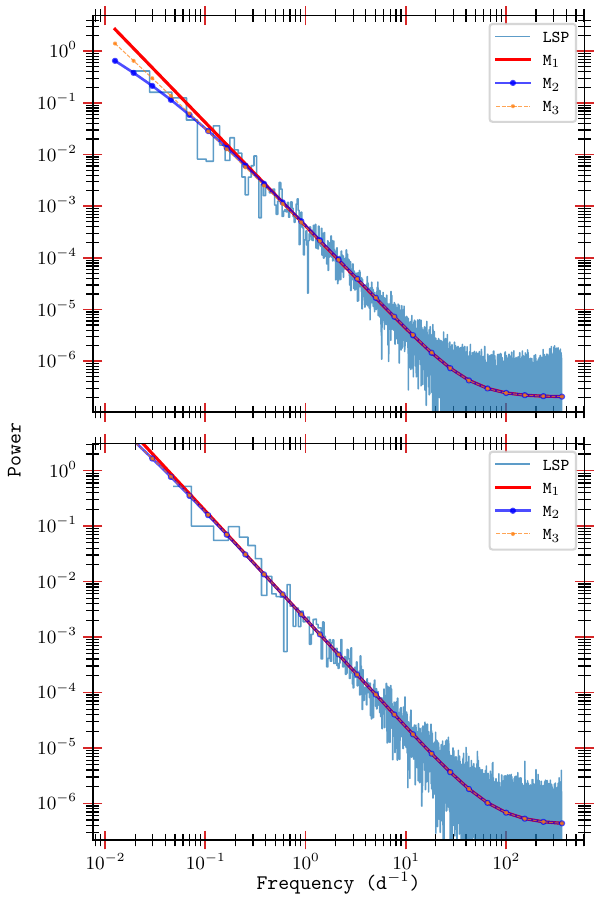}
    \caption{LSP results for rising (upper panel) and the decaying (lower panel) parts}
    \label{fig:LSP}
\end{figure}
\\
\\
It is important to note here that the pixel size for the TESS CCD is large enough to get contaminated by the nearby objects, and the actual variability of the source under probe can be affected by such extraneous contributions. This is also the case for PKS~0735+178, which is accompanied by two nearby faint objects. We tested the variability of these two accompanying objects by inspecting the r-filter data of Zwicky Transient Facility (ZTF), which is within the TESS bandpass filter system. Employing a differential photometry on the source\footnote{\href{https://www.lsw.uni-heidelberg.de/projects/extragalactic/charts/0735+178.html}{field of view of PKS~0735+178}} and its two accompanying objects (at left and right) of the ZTF data, using the comparison object denoted as `D' \citep[chosen because of its less crowded location, see][]{1985AJ.....90.1184S}, we found no obvious variations in the flux values of the two side-objects; hence, any variation observed in the TESS differential LC of the source can be immediately attributed to as source intrinsic variations. Since the comparison object `D' has no nearby contaminating objects so there is no need for such contamination inspection for the `D'. After this basic scrutiny, we generated the aperture photometric LCs of the source and the comparison object along with an estimation of the background flux counts from the $11\times11-$pixels cutouts of the Full-Frame Images centered at the source and the `D', provided by the TESS. The `D' flux values were first background-subtracted and further divided by the sectoral median flux values to obtain a set of normalized flat values. The aperture photometric LC of the source is then subtracted by the background counts, followed by division by the flats to get a differential LC of the source. 
Fig.~\ref{fig:dlc} in the Appendix represents the obtained differential LCs along with the background flux, normalized flat (using `D'), and raw flux of the source for each of the three sectors. We find that though the SAP\_FLUX, PDCSAP\_FLUX, and their processed version can comprehend micro-variability patterns, the trends for the short-term variability analysis seen in the obtained differential LCs are best depicted by the SAP\_FLUX only. Fig.~\ref{fig:complc} highlights the comparisons of the LCs obtained differently, with the differential LCs for the three sectors. Previously, too, the SAP\_FLUX has been seen to be equally useful for variability studies \citep[e.g.,][]{2021MNRAS.501.1100R}. Hence, we directly utilized the SAP\_FLUX without any further detrending. However, in our analysis, all the LCs were made free of local outliers exceeding three times the local 1-day standard deviation. The source PKS~0735+178 has been observed by TESS in five sectors: 44, 45, 46, 71, and 72, out of which {\tt LightCurveFiles} were available only for sectors: 44, 45, and 46, which include the SAP\_FLUX and the PDCSAP\_FLUX. The current study focuses on the LCs obtained in these three sectors only, which cover the peculiar neutrino event. Fig.~\ref{fig:complete_LC} presents the composite LC of the three sectors thus obtained.
\section{Flux Distribution}
\noindent
The composite LC, as shown in Fig.~\ref{fig:complete_LC}, depicts a strong flare event spanning around 75~days. For simplicity here, we have considered the starting epoch of the flare to be $\sim$BTJD~2504.3, where the flux values reach a global minimum in the complete LC ($\sim$120~$e^-s^{-1}$). Visual inspection of the LC indicated that the object undergoes a rapid flux increase, reaching a maximum count of $\sim$375 $e^-s^{-1}$ at $\sim$BTJD 2558.1, followed by an even faster decay. 
Flux distribution is one of the defining properties of accretion-powered sources, as AGNs, often showing normal or log-normal distribution \citep[e.g.,][and references therein]{2016ApJ...822L..13K,2017ApJ...849..138K}. 
The observed flux distribution can be interpreted within the standard framework of variability in accreting systems, where log-normal flux distributions and linear rms–flux relations arise naturally from multiplicative processes in the accretion flow. This behavior has been extensively established in X-ray binaries and AGN \citep[e.g.,][]{2002MNRAS.332..231U, 2005MNRAS.359..345U}, and is often taken as evidence for inwardly propagating fluctuations in the disk. In this context, the flux distribution of PKS 0735+178 is consistent with stochastic variability driven by such multiplicative processes.
The symmetric/asymmetric nature of the profile can be utilized to find the dominance of the cooling process during the LC variations \citep{1999MNRAS.306..551C}. For a well-resolved flare, we also expect the distribution of flux to follow a normal or log-normal nature. 
Assuming that the flare spans from $\sim$BTJD~2504.3 to $\sim$BTJD~2578.7, we evaluated the flux distribution between these epochs of the 2-min cadence LC and present the estimated profile in Fig.~\ref{fig:flux_dist}. The profile has been found to be uneven, so we also evaluated the distribution for rising and decaying segments individually (see Fig.~\ref{fig:flux_dist}). We find that the rising part consists of almost well-separated (equidistantly located within uncertainty) distinct Gaussian profiles, and the unevenness in the profile for the complete LC is mainly due to that from the decaying part. We also note that the high flux values (vicinity of $\sim$350 $e^-s^{-1}$) mainly become part of the decaying portion of the LC. This is highly likely due to the selection bias introduced by the partition of the LC into rising and decaying regions, which was performed via visual inspection.
\section{Excess variance}
\noindent
Accompanied by an additional variability because of the instrumental measurement errors in the observations, the typical variance test can lead to erroneous results. Since flux variability acts as one of the elementary properties of a typical blazar across all EM bands, incorporating the intrinsic measurement uncertainties becomes crucial.  The excess variance method assertively takes care of this uncertainty in quantifying the variability. Given as 
\begin{equation}
    F_{var}=\sqrt{\frac{S^2-\overline{\sigma_{err}^2}}{\bar{x}^2}}\ ,
\end{equation}
\begin{equation}
\&\ \ \ (F_{var})_{err}=\sqrt{\left[\sqrt{\frac{1}{2n}}\frac{\overline{\sigma_{err}^2}}{F_{var}\bar{x}^2}\right]^2+\left[\sqrt{\frac{\overline{\sigma_{err}^2}}{n}}\frac{1}{\bar{x}}\right]^2},
\end{equation}
the fractional root-mean-square variability amplitude ($F_{var}$), or the square root of normalized excess variance, and its corresponding uncertainty, have been computed as in \citet{2003MNRAS.345.1271V}. Here, $S^2-\overline{\sigma_{err}^2}$, $~\bar{x},~\text{and}~n$ are the uncertainty corrected variance, mean flux, and the number of data-points, respectively.
Over the span from BTJD~2504.3 -- 2578.7 of the 2-min cadence data, the average flux is $209.23~e^-s^{-1}$, and the source exhibits an overall variability amplitude of $25.16\pm0.01\%$. The very small associated uncertainty marks the statistical errors only and does not include the realization scatter intrinsic to a single stochastic LC segment.
\section{Variability Timescale}
\noindent
As in \cite{2024ApJ...960...11K}, we utilized the halving/doubling time approach to estimate the variability timescale of the rising ($\sim$BTJD~2504.3 -- 2558.1) and decaying ($\sim$BTJD~2558.1 -- 2578.7) portions individually, expecting them to follow a normal Gaussian distribution. We put similar selection criteria for timescale estimation, choosing only those data point pairs whose flux difference was three times more than the corresponding uncertainty. As the number of such eligible pairs behaves like $n^2$ ($n$ being the number of data points), we note that a 30-min (rather than 2-min) binned LC of the object was utilized to reduce the computational strain due to an extremely large number of eligible pairs for timescale estimation.  The halving/doubling timescales are given as 
\begin{equation}
    F(t)=F(t_0)\*2^{(t-t_0)/\tau}\ \ \ \ (t>t_0) ,
\end{equation}
Fig.~\ref{fig:doub_ts} includes the doubling and halving timescales of the rising and decaying portions of the flare, respectively. We find that the obtained distributions have a major contribution from a single timescale, but unlike in \cite{2024ApJ...960...11K}, they cannot be completely satisfied by single Gaussian functions, showing a multi-modal profile. The most probable timescales dictating the flux variations for the two regions are found to be $43\pm6$~days and $17\pm3$~days, which simply indicates a much faster decay rate than the rising.
\section{Periodogram Analysis}
\begin{table}[]
    \caption{LSP fitting parameters for the rising and decaying epochs of the flare; the uncertainties represent the maximum of upper and lower uncertainty values. Note: BIC criteria favour the rising and decaying epoch to be characterized by $M_1$ (SPL).}
    \hspace{-1.0cm} 
    \resizebox{1.05\linewidth}{!}{\begin{tabular}{c c c c} \hline\hline
        Parameters & & Rising & Decaying \\
        \hline
        Norm. ($A$) & & $(4.17\pm0.15)\times10^{-4}$ & $(2.13\pm0.10)\times10^{-3}$\\
        Spec. index ($\alpha_1$)& & $2.00\pm0.01 $ & $1.95\pm0.01$\\
        Offset ($c$) & & $(2.03\pm0.03)\times10^{-7}$ & $(4.15\pm 0.10)\times10^{-7}$\\
        \hline
    \end{tabular}}
    \label{tab:LSP_paramenters}
\end{table}
\noindent
The power spectral density~(PSD) profile offers useful information regarding the source of variations observed in the LC. AGNs (or blazars) often show a power law (PL) PSD profile ($P(\nu)=A\nu^{-\alpha}$) with the value of $\alpha$ within a vicinity of two, but high values to up to 3.3 have also been reported \citep[e.g.,][and references therein]{2011ApJ...743L..12M, 2013ApJ...770...60S, 2014ApJ...785...60R, 2018ApJ...857..141S}; however, other profile structures have also been discovered, including bending power laws (BPL) \citep[e.g.,][and references therein]{2012A&A...544A..80G}. The PSD characteristic timescales have also been observed to be related to the central black hole mass \citep{2006Natur.444..730M}.\\
\\
For PSD analysis of the LC in Fig.~\ref{fig:complete_LC}, we utilized the 2-min cadence LC. The rising and decaying parts of the flare were first made devoid of big data gaps using linear interpolation to make the segments uniform and get rid of spectral leakages. We then implemented the generalized Lomb-Scargle periodogram \citep[LSP, see][and references therein]{2009A&A...496..577Z} individually to both the rising an decaying part of the flare to inspect the variations in either the PSD shape or the spectral index, testing them with three different models \citep[with $\alpha$ as spectral slope, following][]{2012A&A...544A..80G, 2013MNRAS.433..907E}: \\
$M_1$ (Simple Power law -- SPL) \begin{equation}
    P(\nu)=A\nu^{-\alpha_1}+c
\end{equation}
$M_2$ (Bending Power law -- `BPL1')
\begin{equation}
    P(\nu)=A\nu^{-1}\left[1+\left(\frac{\nu}{\nu_b}\right)^{\alpha_1-1}\right]^{-1}+c
\end{equation}
and $M_3$  (Bending Power law -- `BPL2')
\begin{equation}
    P(\nu)=A\nu^{-\alpha_2}\left[1+\left(\frac{\nu}{\nu_b}\right)^{\alpha_1-\alpha_2}\right]^{-1}+c
\end{equation}
where the free parameters $A,~\alpha_1, ~\alpha_2, ~\nu_b~\text{and} ~c$, are the normalization, spectral indices, bending frequency, and an additive constant, respectively. The above three models respectively have three, four, and five free parameters. 
The best model selection describing the PSD was done using minimization of the Bayesian Information Criterion (BIC), which relies on the log-likelihood statistic given as 
\begin{equation}
    BIC=k\cdot\text{ln }(n)~-~\mathcal{L}
\end{equation}
and the log-likelihood statistic is given as 
\begin{equation}
    \label{lglkd}
    \mathcal{L} = -2\sum_{j}\frac{I_j}{P_j}+\text{log }P_j
\end{equation}
where $I_j$ and $P_j$ represent the estimated and the fitted PSD of the LC, respectively. Following \citet{ 2005A&A...431..391V, 2010MNRAS.402..307V}, individual model parameters are estimated by minimizing $S~(\equiv-\mathcal{L})$, and the associated uncertainty limits can be estimated under the condition $\Delta S = S (\theta)-S_{min}\leq 2.71$, corresponding to 90\% confidence limits \citep[see Appendix A of][]{2005A&A...431..391V}, where $\theta$ represents model parameters.\\
\\
Fig.~\ref{fig:LSP} includes the generalized LSP results of the rising and decaying parts of the LC.  Considering the BIC criteria, we find that both the rising and decaying parts can be illustrated with the simple power law model--$M_1$. Table~\ref{tab:LSP_paramenters} includes the LSP fitting parameters with the simple power law, demonstrating a fractional variation in the power law slopes for the rising and decaying parts. Similar to that of fractional variability amplitude, we again note that estimated parameter uncertainties represent statistical or fit errors only, without including the realization scatter owing to the stochastic nature of the LC segment. 

\section{Discussion and Conclusion}
\noindent
In this work, we carried out high-cadence optical LC analysis of the blazar PKS 0735+178, which was observed from TESS in early December 2021 when the neutrino event association by the source was reported \citep[e.g.][and references therein]{2023MNRAS.519.1396S,2023ApJ...954...70A}. The presented data is the densest optical sampling of the flare reported hitherto, temporally coincident with the alerts: Baksan, IceCube,  Baikal, and KM3NeT at $\sim$BTJD~2553.1, 2557.3, 2557.5, and 2563.9, respectively. The main motivation of the present study is to see how the optical flux of the blazar during the neutrino detection epoch compares with the pre- and post-neutrino detections. 
We find a highly resolved enhanced optical flux during the time period of the reported neutrino alerts compared to the pre- and post-alert phases. The former three alerts lie in the rising phase, with IceCube and Baikal alerts very near to the optical maxima ($\sim$BTJD~2558.3), while the alert with the KM3Net appears in the decaying phase of the flare. While such high optical states are a common feature of flaring blazars and do not, by themselves, establish a physical association with neutrino emission, they are consistent with conditions conducive to neutrino production. In particular, an enhanced optical state may reflect increased particle acceleration and enhanced target photon densities within the emission region, both of which are key ingredients in leptohadronic scenarios.
The TESS optical LC of the blazar PKS 0735+178 is displayed in Fig.~\ref{fig:complete_LC}. The total LC covers the duration of BTJD $\sim$~2500 to $\sim$~2580. In this duration the flux of the source has systematically increased from $\sim$~120~$e^{-}s^{-1}$ to $\sim$~370~$e^{-}s^{-1}$ from BTJD $\sim$~2502 to $\sim$~2559, followed by a sharp decrease from $\sim$~370~$e^{-}s^{-1}$ to $\sim$~140~$e^{-}s^{-1}$ from BTJD $\sim$~2559 to $\sim$~2580. 
Multiwavelength investigations for the spectral energy distribution have been carried out previously to infer the emission properties covering $\gamma$-ray to radio bands \citep{2023MNRAS.519.1396S,2023ApJ...954...70A} and possibilities of different models satisfying the observed neutrino events, although they lack a much denser dataset as achieved with the TESS.
A recent study by \cite{2025A&A...699A.381K} of this blazar in the radio regime (15~GHz and 43~GHz) using very long-baseline interferometry found ejection of a new component (C2) from the core, which crossed a comparatively slowly moving component (C1) along the jet. Using linear back-extrapolation, they estimated that ejection of C2 with a superluminal apparent speed of $\sim4.2c$ is $172.7\pm34.5$~days before the epoch of \href{https://gcn.nasa.gov/circulars/events/icecube-211208a}{IceCube-211208A}. With the available TESS data, assuming that ejection occurred at around the minima of the LC, this ejection epoch lies at around 50~days before the IceCube-211208A. The difference in these two ejection epochs may be due to the fact that C2 started with an even higher apparent speed. This assessment can only be vaguely declared due to the lack of TESS observations of this source at earlier epochs.
The multiple detections or alerts of neutrino events are between BTJD $\sim$~2553 and $\sim$~2564, which is apparently the epochs when C2 crossed C1.
This leads to two possibilities for the neutrino emission region. First case suggests that it is produced near to the core with a possibility of a time lag in which optical flux is leading neutrino detection, due to which source flux started increasing, and after detection of the neutrino event, the flux declined sharply or second where it may happen that only during the interaction of C2 with C1, after a certain energy level or plasma density for some radiation process is reached, the neutrino emission is favored. The latter cause is more plausible because of the very narrow time window of the neutrino detection, with epochs coinciding with those of optical maxima.\\
\\
Owing to the simple causality relation, the most probable variability timescales (here, the doubling timescale $\equiv\tau_\mathrm{var}$) estimated for the rising  phase (assuming that C2 is responsible for the flare) of the flare can be utilized to put limits on the corresponding emission region size (R), given as 
\begin{equation}
R \leq \frac{c ~\tau_\mathrm{var} ~\delta}{1+z} ~,
\end{equation}
where $\delta$ is the Doppler factor. Considering $\delta = 7.59\pm3.15~ \text{and}~ z=0.45\pm0.06$ \citep{2025A&A...699A.381K, 2012A&A...547A...1N}, we find it to be $(5.8\pm2.6)\times10^{17}$cm. The composite LC is seen to have a flaring asymmetric profile with a comparatively slower rise and faster decay (SRFD) that describes vital clues about particle acceleration, energy loss, and the physical environment inside relativistic jets. The SRFD profile is often attributed to the contrast between slow, sustained particle acceleration within the jet and rapid radiative energy losses, which can arise from changes in the jet structure, such as variations in the Doppler factor, relativistic magnetic reconnection events, or geometric effects like jet precession \citep[e.g.][]{1999MNRAS.306..551C,2021MNRAS.504..416S,2026A&A...708A.382K}. \\
\\
Although it cannot be uniquely constrained by the current data, and the above alternative explanations are viable, we propose here one of the probable scenarios for the displayed optical behavior, including a system consisting of an emission core near the central black hole capable of injecting plasmas in the magnetized jet, apparent as distinct components (C1 or C2) in the radio images. The C1 propagates with a slower speed along the jet, and at a later stage, C2 is ejected from the core at a higher superluminal speed, which later crosses the C1. The flare observed here seems to be mainly due to the newly injected component C2. 
Visual inspection of the flux distribution of the rising part of the flare suggests that there are several flux states during the rise. In addition, the flux differences between consecutive humps in the distribution are found to be nearly equal (within corresponding uncertainties). Considering the proposed scenario, this simply suggests that injections of nearly a similar amount of plasma from the proposed emission core to the jet are occurring at regular intervals of time, which collectively appear as a single component C2 as it propagates along the jet and reaches the upcoming knot C1. Looking at the doubling timescale distribution of the rising part, two major humps (at $\sim$43~days and $\sim$61~days) can be perceived. In the current scenario, the bigger hump may correspond to the overall flux doubling rate of the rise, while the smaller one appears because plasma gets injected in chunks, after particular intervals, delaying the rise and giving an additional longer rising timescale. As C2 approaches C1, it is possible that disruption of the compact individual components has happened, and the accumulated plasma is fragmented into an ensemble of randomly moving and variably sized fragments. This aligns with our proposed notion in the manner that multiple dissimilar-sized plasma blobs will have their own individual flux states and will be reflected as a very irregularly shaped overall flux distribution as seen for the decaying portion in Fig.~\ref{fig:flux_dist}. The continuous increase in the full-width-half-maxima of C1 and C2 in the radio images of \cite{2025A&A...699A.381K} aligns with this idea of plasma evolution. Also, because these small fragments have now rather random movements, there will be a major decrease in the Doppler factor (because of variation in the viewing angle of the fragmented plasma motion), causing a much sharper flux decay. This would be consistent with the halving timescale of the decay (Fig.~\ref{fig:doub_ts}), which is much less than that of rising. A meager variation in the spectral slope ($\alpha$) suggests that fragmented plasma blobs have a similar particle distribution as that of the parent distribution in the flux frozen state, differing only in their respective sizes.\\
\\
More than two decades ago, it was predicted that some of the blazars may be neutrino loud, and the small sample of potential sources mentioned was dominated by strong TeV-emitting HBL blazars \citep{2002PhRvD..66l3003N}. To date, neutrino emissions have only been reported from four blazars: TXS 0506+056, PKS 1424+240, GB6 J1542+6129, and PKS 0735+178, and most of them do not belong to strong TeV-emitting HBL blazars \citep{2023MNRAS.519.1396S}. In contrast to ``real" BL Lacs, which are intrinsically featureless or very weak emission-lined, \citet{2017A&ARv..25....2P, 2019MNRAS.484L.104P, 2022MNRAS.511.4697P} found that all four neutrino-emitting blazars are masquerading BL Lacs, namely FSRQs, whose emission lines are overwhelmed by an extremely bright, Doppler-boosted jet. As HEGs (high excitation galaxies) with intrinsically strong emission lines dominating their composite optical/UV spectra, masquerading BL Lacs benefit from a number of radiation fields outside the jet, such as the dusty torus, photons reprocessed in the broad-line region (BLR), and the accretion disc, which may increase neutrino production by giving the protons more targets.  
PKS 0735+178 has been discussed as a potential neutrino-emitting blazar in several recent multi-messenger studies. Detailed broadband and leptohadronic modeling efforts \citep[e.g.,][]{2023MNRAS.519.1396S, 2023ApJ...954...70A, 2025ApJ...989..208P} indicate that, while hadronic scenarios can reproduce the observed spectral energy distribution, the predicted neutrino flux is typically low and strongly model-dependent. Even leptonic models can also adequately explain the electromagnetic emission. We note that the temporal proximity of the observed electromagnetic activity to the reported neutrino event may suggest a possible physical connection, potentially within a common hadronic emission framework. However, given the limited statistical significance and the absence of simultaneous multi-messenger constraints, such an association remains tentative. Currently, the sample of neutrino-emitting blazars is very tiny, and we need more time to detect neutrino emission from many more blazars, and then we can make a conclusive statement. Alternative explanations, including a chance coincidence, cannot be excluded. Therefore, the neutrino–emission link should be regarded as suggestive rather than conclusive, pending further observational evidence.\\
\\
{\it Acknowledgment:} We thank the reviewer for very useful comments, which helped us to improve the manuscript. This paper includes data collected with the TESS mission, obtained from the Mikulski Archive for Space Telescopes (MAST) data archive at the Space Telescope Science Institute (STScI). Funding for the TESS mission is provided by the NASA Explorer Program. STScI is operated by the Association of Universities for Research in Astronomy, Inc., under NASA contract NAS5–26555. The specific observations analyzed can be accessed via \dataset[doi:10.17909/q0dp-x410]{https://doi.org/10.17909/q0dp-x410}.
\\\\
{\it Software:} lightkurve \citep{2018ascl.soft12013L}, 
SciPy \citep{2020SciPy-NMeth}, PyAstronomy \citep{pya}
% 10.17909/q0dp-x410
\bibliography{ref} 
\bibliographystyle{aasjournalv7}
\section*{appendix}
\begin{figure*}[h]
\centering
\includegraphics[width=.47\linewidth]{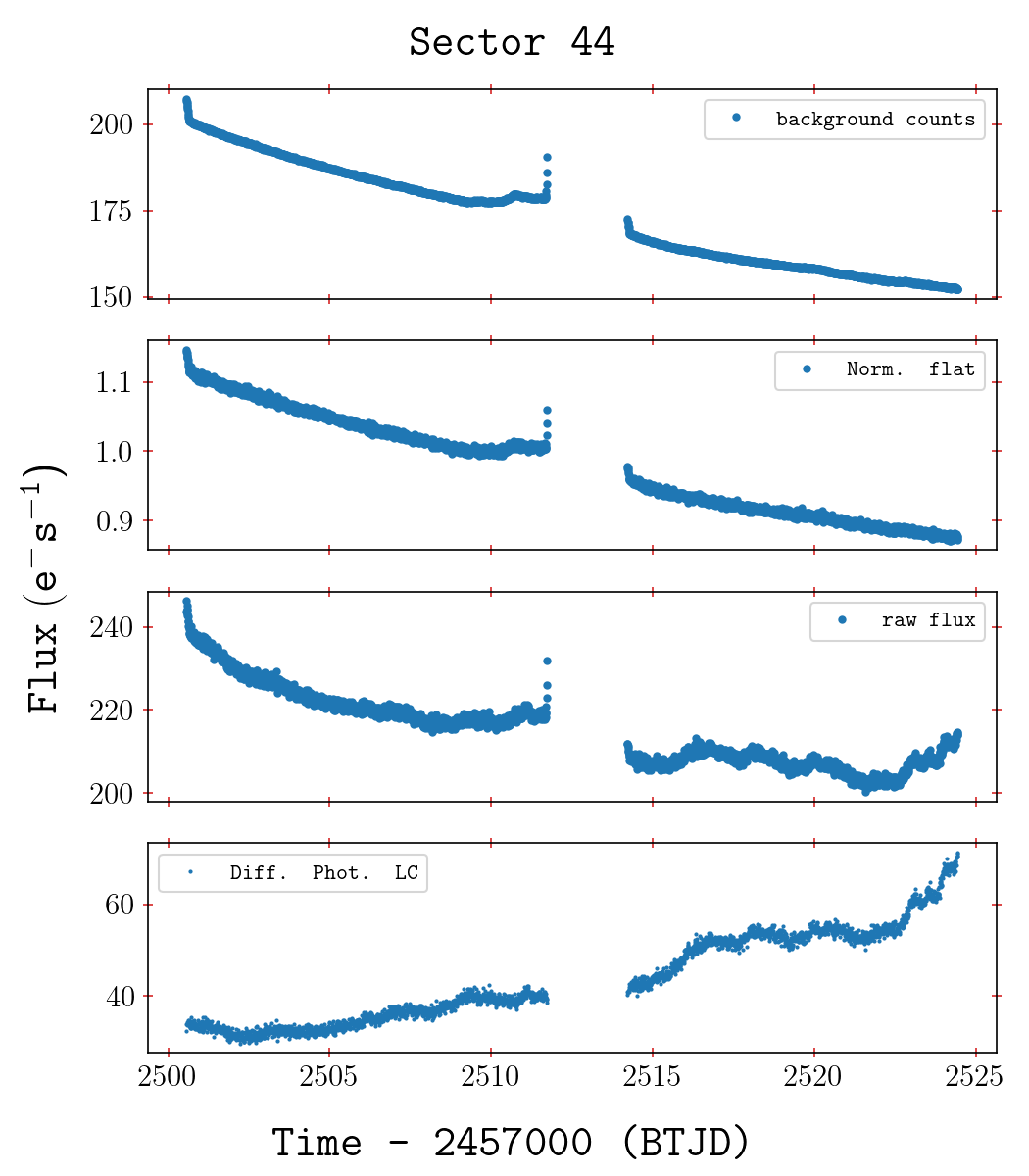}
\includegraphics[width=.47\linewidth]{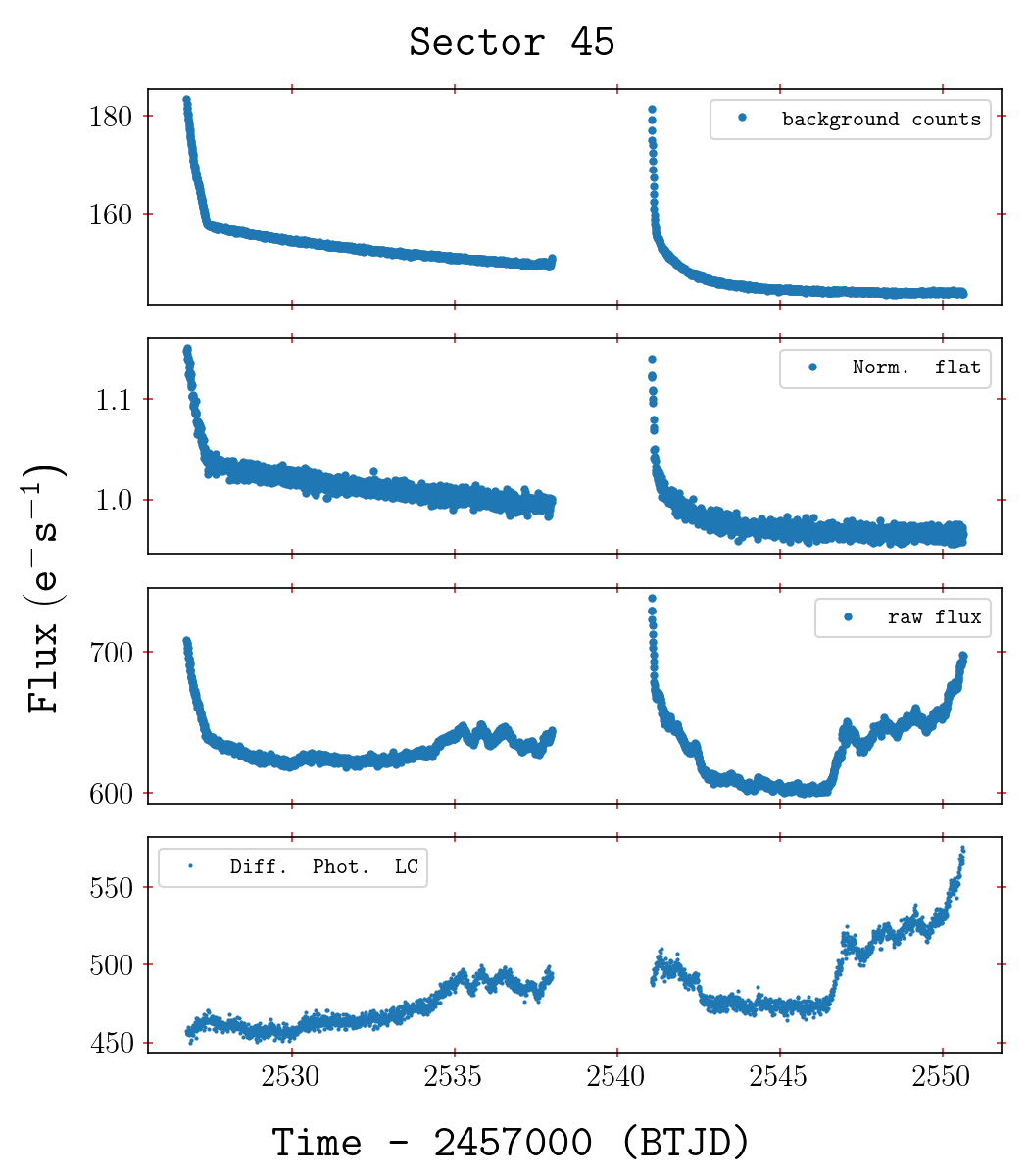}
\includegraphics[width=.47\linewidth]{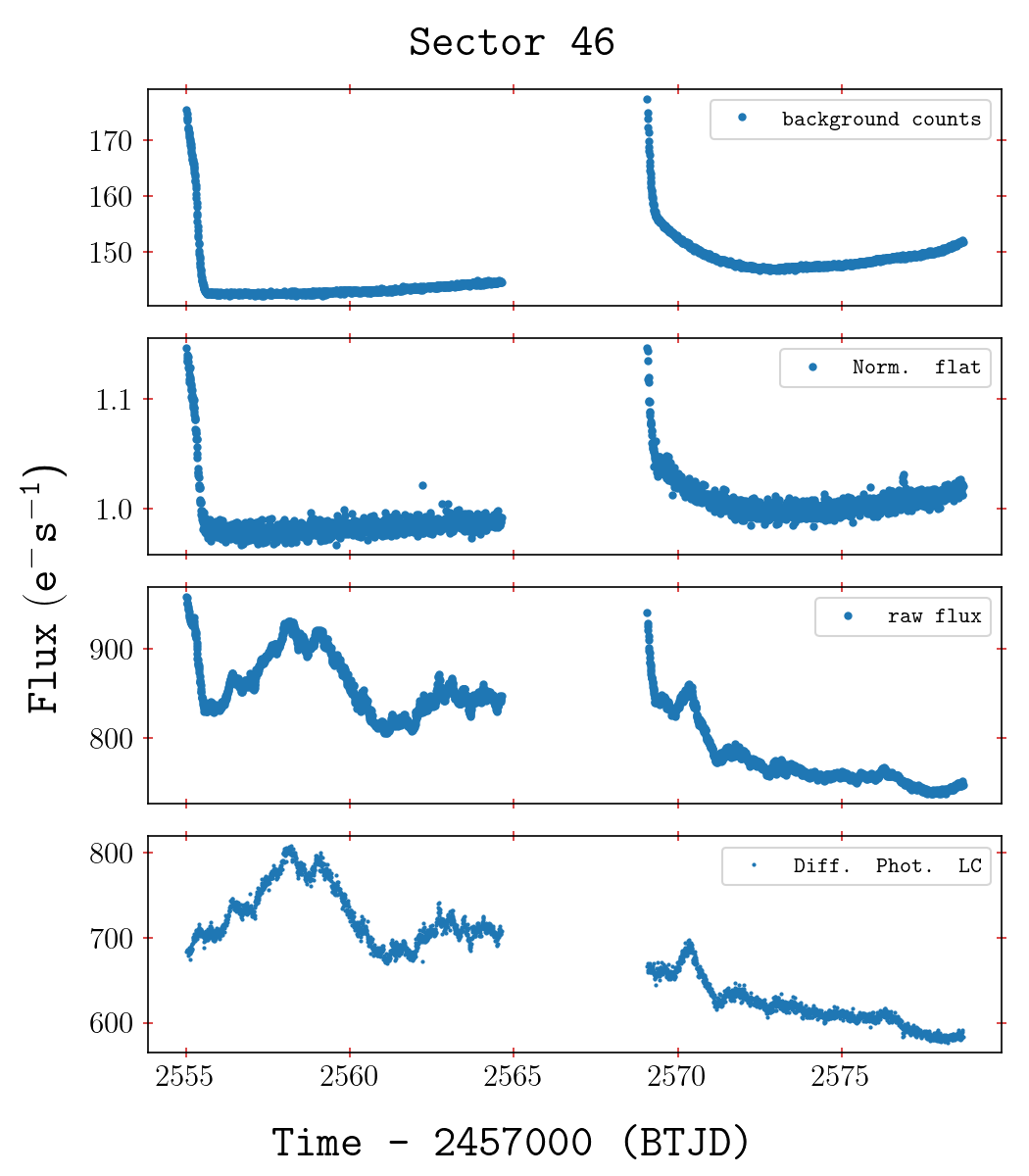}
\caption{Each of the indicated sectors includes the flux counts obtained from the $11\times11$ full frame image cutout. For each sector, the normalized (Norm.) flat (second panel) has been estimated using the comparison object `D' in the image cutout, where the median flux value of `D' is used for normalization. The Norm. flats under 15\% only around unity have been considered for comparison of the differential LC with the LCs obtained with other methods (Fig.~\ref{fig:complc}). The differential photometric LC (Diff. Phot. LC, fourth panel) is then obtained by dividing the raw flux of the source by the Norm. flat.}
\label{fig:dlc}
\end{figure*}
\end{document}